\documentclass[pre,amssymb,reprint,superscriptaddress,showpacs]{revtex4-1}
\usepackage{amsmath}
\usepackage{amsfonts}
\usepackage{amssymb}
\usepackage{epsfig}
\usepackage{graphicx}

\renewcommand{\phi}{\varphi}

\newcommand{\be}{\begin{equation}}
\newcommand{\ee}{\end{equation}}
\newcommand{\bea}{\begin{equnaray}}
\newcommand{\eea}{\end{equnaray}}
\newcommand{\ba}{\begin{align}}
\newcommand{\ea}{\end{align}}

\usepackage{color}

\begin{document}

\title{Macroscopic yielding in jammed solids is accompanied by a 
non-equilibrium first-order transition in particle trajectories}
 
\author{Takeshi Kawasaki}
\affiliation{Laboratoire Charles Coulomb, UMR 5221 CNRS, 
Montpellier, France}
\affiliation{Department of Physics, Nagoya University, Nagoya 464-8602, Japan}

\author{Ludovic Berthier}
\affiliation{Laboratoire Charles Coulomb, UMR 5221 CNRS, 
Montpellier, France}

\date{\today}

\begin{abstract}
We use computer simulations to analyse the yielding transition during
large-amplitude oscillatory shear of a simple model for 
soft jammed solids. Simultaneous analysis of global  mechanical 
response and particle-scale motion demonstrates that 
macroscopic yielding, revealed by a
smooth crossover in mechanical properties, is accompanied by a sudden
change in the particle dynamics, which evolves from non-diffusive motion
to irreversible diffusion as the amplitude of the shear
is increased. We provide numerical evidence that this sharp change 
corresponds to a non-equilibrium first-order dynamic  phase transition, thus
establishing the existence of a well-defined microscopic dynamic signature 
of the yielding transition in amorphous materials in oscillatory shear.
\end{abstract}



\maketitle

\section{Introduction}

A major effort in soft condensed matter physics concerns the 
design of materials with well-controlled mechanical properties~\cite{Larson}. 
Rheology thus represents a central probe and  
oscillatory shear measurements at finite frequency $\omega$ 
are among the most commonly performed mechanical 
tests~\cite{review2011}. 
In this approach, a harmonic deformation 
is applied and the stress response measured, or vice-versa. In the linear 
regime, the complex shear modulus $G^*(\omega) = G'(\omega) 
+ {\rm i} G''(\omega)$
provides information about the nature and strength 
of the material at a given frequency~\cite{olafsen}, while 
microscopic relaxation processes can be probed by
varying the frequency. At larger amplitude, 
non-linear mechanical properties are accessed. 

This approach is well-suited for amorphous materials, 
which often display non-trivial response spectra 
in the linear regime where they behave as soft elastic solids, 
but flow as the amplitude of the forcing
is increased beyond a ``yielding'' limit~\cite{sgr,rmp,bouchbinder}. 
Whereas the storage modulus $G'(\omega)$ 
dominates the elastic response at small deformation
amplitude, irreversible plastic deformations occur post yielding 
where the loss modulus $G''(\omega)$ instead dominates.
Oscillatory shear experiments have been performed in a wide range of 
soft condensed matter systems across yielding,  
such as granular particles~\cite{pine2005,pine2008,keim,keim2,hecke}, 
emulsions~\cite{luca,emulsion,Munch_emulsion,weeks}, 
colloidal suspensions~\cite{pusey,george,ganapathy} and 
gels~\cite{rogers}, as well as in  
computer simulations~\cite{mohan,fuchs,priez,Richhardt,foffi}.

In experiments, the change from elastic to plastic response in macroscopic 
mechanical properties is often described as a ``yielding transition'',
even though yielding appears as a smooth crossover whose location 
cannot be unambiguously defined~\cite{rmp2}. Interestingly,
recent experiments have provided evidence that this macroscopic
crossover corresponds to a qualitative change in particle 
trajectories~\cite{hecke,keim,luca,ganapathy,schall2,schall,regev}. 
As expected physically, particles are essentially arrested in
the undeformed solid, but can diffuse due to 
irreversible plastic rearrangements occurring at larger 
amplitude. There is, however, no 
consensus about the nature of this crossover, which has been
described either as a smooth change \cite{keim2}, as
a relatively sharp crossover \cite{luca}, 
or as a continuous non-equilibrium phase 
transition~\cite{ganapathy}. The latter conclusion builds a
qualitative analogy with the continuous irreversibility transition 
observed in low-density suspensions~\cite{pine2005,pine2008,ohern2}, 
which has been actively studied in computer 
simulations~\cite{Richhardt,foffi}, and attempts
to borrow concepts from the field of non-equilibrium phase 
transitions~\cite{kuramoto_review,hinrichsen,henkel}.  
In addition recent experiments argue that yielding corresponds to a change 
in the microstructure 
of the system~\cite{schall}, by opposition to the dynamic properties
discussed here. 
A clear connection between these microscopic 
changes and the macroscopic rheology is lacking.

Here, we use a model of a 
jammed material composed of non-Brownian repulsive spheres 
to investigate the nature of the yielding transition at the 
particle-scale level, in the simple situation where thermal
fluctuations and hydrodynamic forces play no role. We reproduce 
standard mechanical signatures of macroscopic 
yielding under oscillatory shear and obtain two key results
regarding particle trajectories. 
First, we show that the onset of particle diffusion in steady state
is extremely sharp and occurs at a well-defined shear amplitude, 
which unambiguously locates the  
yielding transition. Second, we find that particle diffusivity 
emerges discontinuously at yielding,
thus demonstrating that yielding corresponds to a
non-equilibrium first-order phase transition. These findings differ 
qualitatively from earlier suggestions of a continuous irreversibility
transition~\cite{foffi,ganapathy}, but seem to agree very well with 
recent experimental findings~\cite{luca,keim2}. We also 
show that this transition is dynamic in nature, and is not 
accompanied by discontinuous structural changes. 

\section{Model and numerical techniques}

We consider soft repulsive non-Brownian particles in a simple shear
flow geometry. We perform standard
overdamped Langevin dynamics simulations of a well-known
model of harmonic particles in three dimensions~\cite{durian,allen}, using an 
equimolar binary mixture of small and large particles with diameter 
ratio 1.4. 
The equations of motion read
\be 
\xi_s \left[ \frac{\partial \vec{r}_i}{\partial t}  - 
\dot{\gamma}(t)y_i(t)\vec{e}_x \right] +
\sum_j \frac{\partial U( r_{ij} )}{\partial 
\vec{r}_{ij}}  = \vec{0},
\label{em}
\ee
where $\xi_s$ is a friction coefficient, 
$\vec{r}_{ij}=(x_{ij}, y_{ij}, z_{ij}) = (x_j-x_i,y_j-y_i,z_j-z_i)$, 
$\vec{e}_x=(1,0,0)$ and $\dot\gamma(t)$ is the shear rate.
For particles $i$ and $j$ having 
diameters $a_i$ and $a_j$, the pair potential reads
$U(r_{ij})=\frac{\epsilon}{2} \left(1-{r_{ij}}/{a_{ij}}\right)^2\Theta 
(a_{ij}-r_{ij})$, where $\epsilon$ is an energy scale, 
$a_{ij}=(a_i+a_j)/2$, $a$ denotes the diameter of small 
particles, and $\Theta (x)$ is the Heaviside function. 
The unit length is $a$, the unit time $\tau_0 = 
a^2 \xi_s / \epsilon$, the unit energy $\epsilon$, and so the
unit stress is $\epsilon/a^3$.

We apply a harmonic deformation using Lees-Edwards periodic 
boundary conditions~\cite{allen}, the strain evolving 
as ${\gamma}(t)= \gamma_0 [1- \cos(\omega t) ]$,
where $\gamma_0$ is the amplitude of the imposed shear strain 
and $\omega = 2 \pi / T$ is the frequency of the oscillation.  
The period $T$ is chosen very large to be close to a
quasi-static protocol; we 
use $T=10^4 \tau_0$. We have checked that our results 
do not qualitatively depend on this choice. 
We work at constant packing fraction $\varphi = 0.80$, 
much above the jamming density
$\varphi_J \simeq 0.647$~\cite{kawasaki15}. We checked 
that our results are representative of the entire jammed phase, 
$\phi > \phi_J$, but yielding could be more complicated in the limit 
$\phi \to \phi_J$ where the system looses rigidity~\cite{luca}.
The different regime $\phi < \phi_J$, where a yield stress does not exist,
was analysed before~\cite{ohern2}.
We solve Eq.~(\ref{em}) with a modified Euler algorithm~\cite{allen} 
using a discretization timestep $\Delta t=0.1\tau_0$.
Numerical stability and accuracy were carefully checked.  
To investigate finite-size effects, we perform simulations with
{four} different sizes $N=300$, $1000$, 3000, and 10000, where 
$N$ is the total number of particles. 
All simulations start from fully random configurations. 
We analyse both the transient regime after shear is started, and 
steady-state measurements.
To improve the statistics, we perform 
at least 4 independent runs for each pair $(\gamma_0, N)$. 

At the macroscopic level, our main observable 
is the time-dependent response of the $xy$-component of the 
shear stress, defined by the usual Irving-Kirkwood 
formula~\cite{allen}:
$\sigma (t) = - \frac{1}{V} \sum x_{ij} F_{ij}^y$,
where $V$ is the volume and $F_{ij}^y$ represents the 
$y$-component of the force $F_{ij}$. The kinetic part
of the stress is fully negligible in the present 
situation of low-frequency oscillatory shear. To analyse
the rheological response in steady state, we fit the 
time series of the shear stress to a sinusoidal form, 
\be
\sigma (t) = -\sigma_0 \cos{(\omega t+\delta)},
\label{sigma_cos}
\ee
where $\sigma_0$ is the amplitude of the first harmonics
at frequency $\omega$, and $\delta$ is the phase 
difference between strain and stress. In practice, 
$\sigma_0$ and $\delta$ are obtained by fitting Eq.~(\ref{sigma_cos})
to steady state data lasting about $100T$.
Alternatively, we can transform the two parameters
$(\sigma_0, \delta)$ into the more conventional quantities
$G'(\omega)$ and $G''(\omega)$ using
\be
G'(\omega)+{\rm i} G''(\omega) = \frac{\sigma_0}{\gamma_0}
e^{{\rm i}\delta}.
\label{gprime}
\ee 

\begin{figure}
\psfig{file=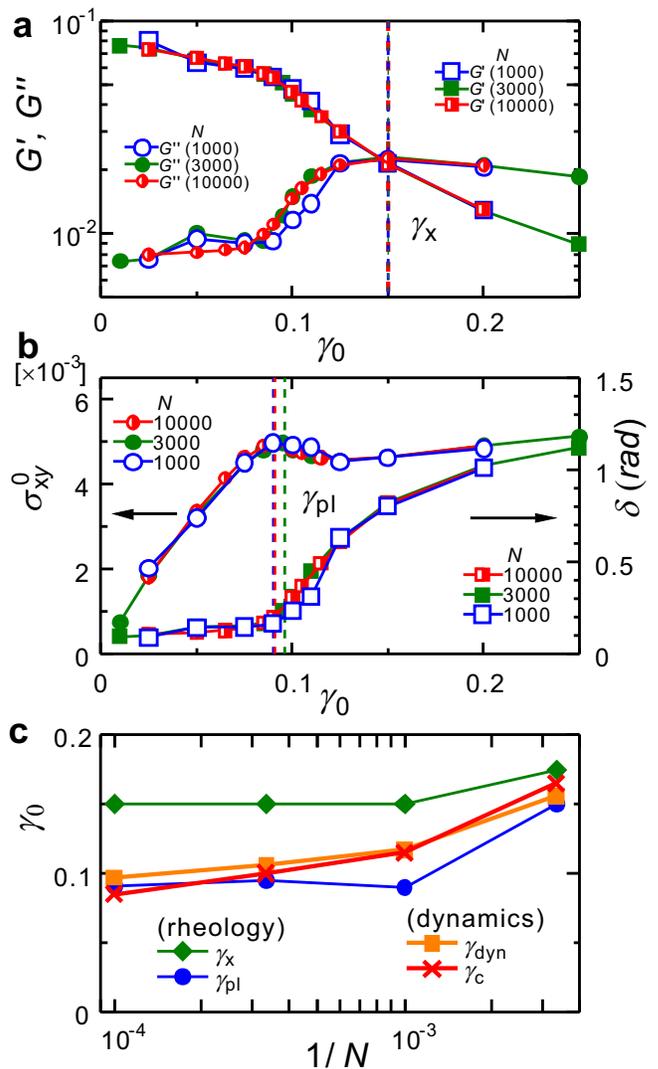,width=8.5cm,clip}
\caption{\label{fig1} 
(a) Evolution of storage and loss moduli, Eq.~(\ref{gprime}), with strain 
amplitude for different system sizes. The crossing 
of $G'$ and $G''$ at $\gamma_{\times}$ (dashed lines) defines a
characteristic strain scale.
(b) Evolution of amplitude and phase of the stress, 
Eq.~(\ref{sigma_cos}), with strain amplitude. The amplitude 
has a maximum at $\gamma_{\rm pl}$ (dashed lines).
(c) System size dependence of all characteristic strain amplitudes.}
\end{figure}

\section{Smooth crossover in macroscopic rheology}

In Fig.~\ref{fig1}(a), we show the evolution with the strain 
amplitude $\gamma_0$ of the storage and loss moduli at fixed 
frequency $\omega$ measured in steady state.
At very low $\gamma_0$, $G'$ dominates the response, 
$G'/G'' \approx 10$, indicating that the system responds in 
the linear regime as a soft elastic solid. As $\gamma_0$ is increased,
the moduli first evolve slowly for $\gamma_0 < 0.1$,
where little plastic rearrangements are produced. 
As $\gamma_0$ increases further, we observe a crossing 
of $G'$ and $G''$ at $\gamma_{\times} \approx 0.15$ (dashed lines), 
so that dissipation dominates 
$\gamma_0 > \gamma_{\times}$. 
These mechanical properties reproduce well-known 
behaviour~\cite{Larson,review2011} and validate our numerical approach.  
We notice further that they display virtually no finite-size effects.
In this representation, $\gamma_{\times} \approx 0.15$ appears as the most 
relevant strain scale to characterize yielding, although a smooth crossover 
can be qualitatively detected near $\gamma_0 \approx 0.1$, 
where the $\gamma_0$-dependence of the moduli becomes somewhat steeper. 

In Fig.~\ref{fig1}(b) we replot this rheological evolution
in the alternative representation offered 
by $(\sigma_0, \delta)$. At small $\gamma_0$, the phase $\delta$ is 
very small while the stress amplitude increases 
linearly, $\sigma_0 \approx G' \gamma_0$, as expected for 
reversible elastic deformations in a solid. 
Near $\gamma_0 \approx 0.1$ two changes are observed. 
First, $\sigma_0$ ceases to be linear and displays an 
overshoot when $\gamma_0$ is increased, signaling that 
plastic events take place.
We define the onset of plastic 
events, $\gamma_{\rm pl}$, as the location of the stress overshoot
[dashed lines in Fig.~\ref{fig1}(b)]. Our interpretation for $\gamma_{\rm pl}$
is reinforced by the evolution of the phase $\delta$
in Fig.~\ref{fig1}(b), which grows steadily above
$\gamma_{\rm pl}$, indicating the onset of dissipation.
Note that the crossing of $G'$ and $G''$ 
at $\gamma_{\times}$ has no obvious relevance in 
this representation where it simply corresponds, by definition, 
to the strain scale where $\delta = \pi/4$. In Fig.~\ref{fig1}(c) 
we confirm that $\gamma_{\rm pl}$ and $\gamma_{\times}$ display 
virtually no system size dependence, but that they differ quantitatively.

Whereas $\gamma_{\times}$ is frequently quoted as ``the'' yielding 
point in the literature, a stress overshoot  
also serves to identify yielding in shear-start experiments~\cite{rmp2}.
A stress overshoot is reported in some oscillatory shear
experiments~\cite{rogers}, 
but is absent in others~\cite{luca,emulsion}. A possible explanation is that 
experiments are typically performed at somewhat larger frequencies, 
where additional contributions to the shear stress 
(lubrication forces, hydrodynamic effects) might hide this behaviour.
In addition, we have confirmed that the overshoot disappears and 
is replaced by a monotonic increase seen in experiments~\cite{emulsion} 
when we use a substantially larger frequency, typically $\omega > 10^{-3}$.
We emphasize that the presence of the 
dynamic transition discussed below is independent of the existence 
of the stress overshoot reported in Fig.~\ref{fig1}(b).

\begin{figure}
\psfig{file=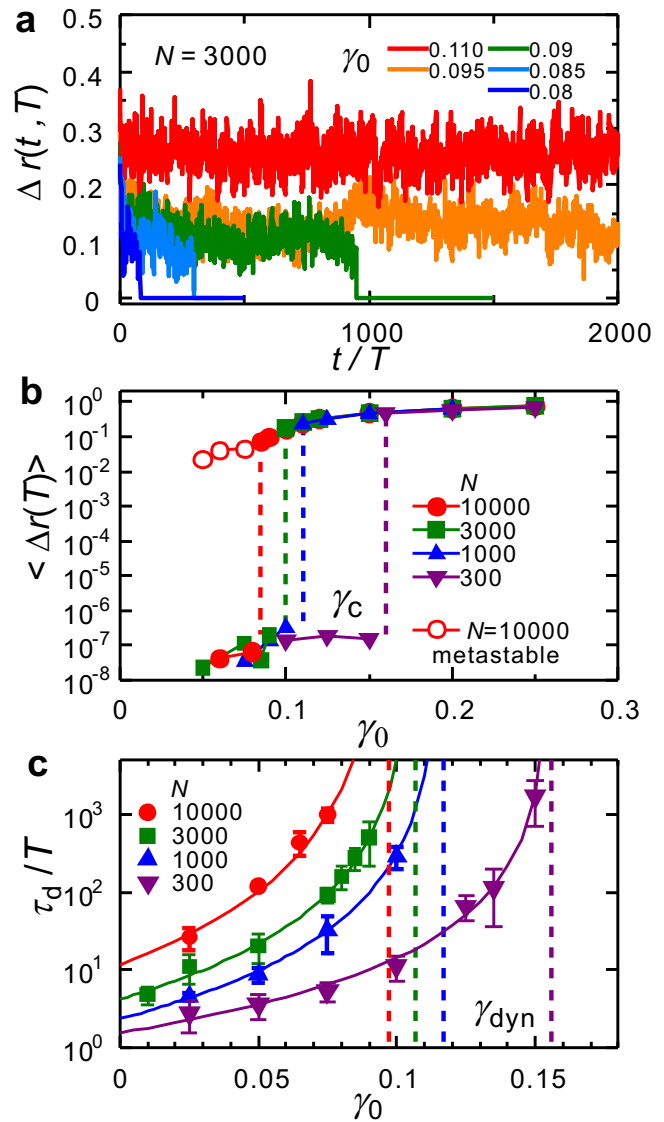,width=8.5cm}
\caption{\label{fig2} 
(a) Transient evolution of particle displacement
$\Delta r(t,T)$, Eq.~(\ref{Delta}), for various 
strain amplitudes for $N=3000$. For $\gamma_0 \lesssim 0.090$, 
the displacements drop to zero after a timescale $\tau_{\rm d}$. 
{
(b) The time averaged displacement for one cycle $\langle\Delta r(T)\rangle$ 
for 
various strain amplitudes and system sizes. 
The stable and the metastable data are plotted with 
filled and open symbols respectively. 
The vertical dashed lines indicate $\gamma_c$ for each system size.
(c)} The average $\tau_{\rm d}$ 
diverges algebraically, $\tau_{\rm d} \sim (\gamma_{\rm dyn}-\gamma_0)^{-\alpha}$,
interpreted as the diverging lifetime of a metastable state near a 
discontinuous phase transition.
{ The vertical dashed lines indicate $\gamma_{\rm dyn}$ 
for each system size.}
}
\end{figure}

\section{Sharp transition in microscopic dynamics}

We now turn to the evolution of single particle dynamics. 
A first natural dynamic observable is the averaged
particle displacement after one deformation cycle~\cite{luca,foffi}, 
\be
\Delta r(t, T) = \frac{1}{N} \sum_{j}
|\vec{r}_j(t+T)-\vec{r}_j(t)|,
\label{Delta}
\ee
where $t$ is the time since shear is applied.
In Fig.~\ref{fig2}(a), we show how $\Delta r(t,T)$ 
evolves in the transient regime for various 
amplitudes of the applied deformation. For small
$\gamma_0$, particle displacements decay rapidly to zero. 
In the elastic solid at small amplitude, there are rare
rearrangements taking place before the system settles 
near a stable energy minimum where particles have nearly periodic motion
(or quasi-periodic motion with a period that is 
a multiple of $T$, as reported before \cite{Richhardt,ohern1}).  
As $\gamma_0$ is increased, it takes more and more time 
for $\Delta r(t,T)$ to eventually vanish. 
When $\gamma_0$ is larger than  $\gamma \sim 0.095$,
the average particle displacement never vanishes 
in the explored time window, but instead fluctuates around 
a well-defined finite value, which increases with $\gamma_0$. 
This regime corresponds to irreversible, non-periodic particle trajectories.
{In Fig.~\ref{fig2}(b), we plot the time averaged 
displacement for one cycle $\langle\Delta r(T)\rangle$ in steady states for 
various strain amplitudes and system sizes. 
From $\langle\Delta r(T)\rangle$, a very clear discontinuous jump 
is observed between the irreversible and reversible states
near $\gamma_{\rm c}$. Very close to the transition, 
the displacements exhibit fluctuations around a 
well-defined value both above and below $\gamma_c$. Whereas 
these fluctuations are infinitely long-lived above $\gamma_c$, they
are only metastable below $\gamma_c$, before the 
system finds a reversible state where the displacements
become very small. We report 
the value of $\langle\Delta r(T)\rangle$ for $N=10000$ 
for these metastable states in Fig.~\ref{fig2}(b).}
Overall, these  fluctuations appear qualitatively distinct from the algebraic 
decay observed close to continuous irreversibility 
transitions \cite{pine2008}, and are much closer to the phenomenology 
observed near discontinuous, first-order phase transitions
where metastable phases can be observed over long times. 
{In particular, it appears impossible to describe the 
decrease of $\langle \Delta r(T) \rangle$ with a continuous 
vanishing at the critical value of $\gamma_{\rm dyn}$.}

Stronger evidence of such a phase transition is obtained from 
the evolution of the average lifetime 
of the metastable irreversible phase $\tau_{\rm d}$, 
as depicted in Fig.~\ref{fig2}(c) for various system sizes.  
These data confirm that $\tau_{\rm d}$ increases rapidly 
close to $\gamma_0 \approx 0.1$. For larger 
$\gamma_0$, trajectories remain irreversible. By contrast to the rheology, 
a clear system size dependence is observed, 
larger systems take more time to settle in 
a global energy minimum. A diverging lifetime 
is typically observed close to non-equilibrium phase transitions 
\cite{pine2008,kuramoto_review,hinrichsen}, and 
was reported before~\cite{Richhardt,foffi}.
Such divergence is expected for both a continuous or discontinuous
transition (see \cite{fiore,xu} for recent examples). 
Solid lines in Fig.~\ref{fig2}(c) represent an empirical 
fit, $\tau_{\rm d} \sim (\gamma_{\rm dyn}-\gamma_0)^{-\alpha}$,
with {$\alpha \approx 2.1-3.0$ \cite{exponent}}, 
suggestive of a divergence of $\tau_{\rm d}$ when approaching the
dynamic transition at $\gamma_{\rm dyn}$. {We notice that 
finite size effects can be felt even when the system is not very close
to the critical value, such that we cannot observe the expected 
saturation of $\tau_d$ to a finite value as $N\to\infty$.}
The system size
dependence of $\gamma_{\rm dyn}$ reported in Fig.~\ref{fig1}(a)
is very modest and seems to extrapolate to a finite value,
$\gamma_{\rm dyn} \approx 0.095$, as $N \to \infty$.  

\begin{figure}
\psfig{file=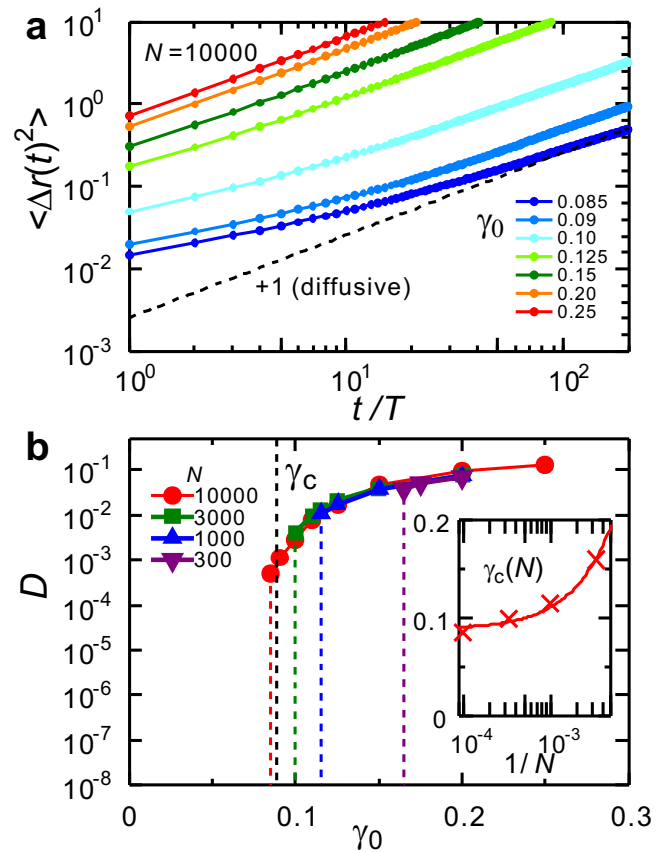,width=8.5cm}
\caption{\label{fig3} 
(a) Mean-squared displacements, Eq.~(\ref{msd}), 
for $N=10^4$ and various strain amplitude become 
diffusive at long time when $\gamma_0 \geq \gamma_c \approx 0.085$. 
(b) The diffusion constant $D$ decreases modestly as $\gamma_0$ 
decreases towards a critical value $\gamma_c$ where it drops 
discontinuously to zero. {The vertical dashed lines indicate 
$\gamma_c$ for each system size and the black one 
represents the large-$N$ extrapolation of $\gamma_{\rm c} 
\approx 0.0885$ performed in the inset.}}
\end{figure}

We now characterize the steady-state irreversible dynamics at large 
$\gamma_0$ using the mean-squared displacement, 
\be
\langle \Delta r(t)^2\rangle= \frac{1}{N} \langle 
\sum_{j=1}^N
|\vec{r}_j(t)-\vec{r}_j(0)|^2 \rangle,
\label{msd}
\ee
where the brackets indicate a time average. The results are displayed 
in Fig.~\ref{fig3}(a) for $N=10^4$, for time delays commensurate with the 
period. The
dynamics is diffusive at long times,
$\langle \Delta r(t)^2\rangle \sim 6Dt$,
where $D$ is the diffusion constant. We represent $D$ 
(in units of $a^2/T$) for various $\gamma_0$ and $N$ in Fig.~\ref{fig3}(b).
As expected, $D=0$ below a critical value $\gamma_{\rm c}$, 
corresponding to the phase characterized by quasi-periodic particle 
trajectories, and it increases with $\gamma_0$ 
above $\gamma_{\rm c}$. 
Both $D$ and $\Delta r(t,T)$ in Eq.~(\ref{Delta}) could serve as order
parameters for the transition.
By measuring $\gamma_c$ for various system sizes, 
we observe a modest change with system size, see 
inset of Fig.~\ref{fig3}(b), suggestive of a finite limit 
 $\gamma_c \approx 0.0885$ for $N \to \infty$. The functional form of our  
extrapolation should be confirmed by additional larger scale 
simulations. 

A striking finding in Fig.~\ref{fig3}(b) is the finite 
amplitude of the diffusion constant 
at the transition. Near continuous irreversibility transitions, 
$D$ decreases by several orders of magnitude and scales
algebraically as $\gamma_c$ is approached from above~\cite{pine2005,elsen}. 
We observe instead a modest decrease of $D$, followed
by a sudden jump to zero, which is robust against 
finite-size effects. {In particular we find that 
diffusive behaviour also persists for a
finite amount of time below in the reversible 
phase, as also described above for the one-cycle particle
displacement $\Delta(r,T)$. It is however more difficult 
to measure $D$ in this region, because a careful determination
of $D$ requires taking the long-time limit, which is not possible
by construction in the metastable region. We conclude therefore
that the discontinuous
behaviour of $D$ in Fig.~\ref{fig3}(b) appears less convincing than 
the one of $\Delta (r,T)$ shown in Fig.~\ref{fig2}(b), but the overall
phenomenology reported in this work appears inconsistent with a continuous
transition.}

\section{No change in microscopic structure}

It was recently argued that the yielding transition in oscillatory shear 
can be detected through the static structure of the system~\cite{schall}. 
Such a behaviour would differ qualitatively from our conclusion
that yielding is revealed through the dynamic evolution of the
system. 
Our analysis of the pair correlation function across the yielding 
transition did not reveal any change in the static properties 
of the system in the two phases, which seems to contradict
the results of Ref.~\cite{schall}. To reinforce this conclusion, 
we have measured the exact same quantity that was detected experimentally.
In detail, we resolve the radial dependence of the static structure factor
$S(\vec{q})= \frac{1}{N} \langle \rho(\vec{q})\rho(-\vec{q}) \rangle$ 
in the $(x,z)$ plane \cite{schall}, where $\rho(\vec{q})$ is the Fourier
component of the density at wavevector $\vec{q}$.
Thus we define the angle $\alpha$ from $\tan \alpha = {(q_z/q_x)}$,
and follow the $\alpha$-dependence of the static structure, 
as proposed in \cite{schall}.

To obtain statistically reliable data close to the dynamic transition, 
we perform an extensive time average over 100 well separated 
times for the diffusive phase at $\gamma_0=0.12$. 
For the non-diffusive phase at $\gamma_0=0.10$, time averages
are not useful and we obtain instead 100 independent configurations 
starting from independent initial conditions.
Errorbars are defined from the 
resulting sample-to-sample fluctuations. 
Because the stress is close to a sinusoidal form, we measure 
the structure either when the stress is zero (`undeformed' states)
and when it is maximal $\sigma(t) = \pm\sigma_0$ (`deformed' states). 
Thus, we obtain 4 measures of the structure at two shear amplitudes,
for both deformed and undeformed states. 

\begin{figure}
\psfig{file=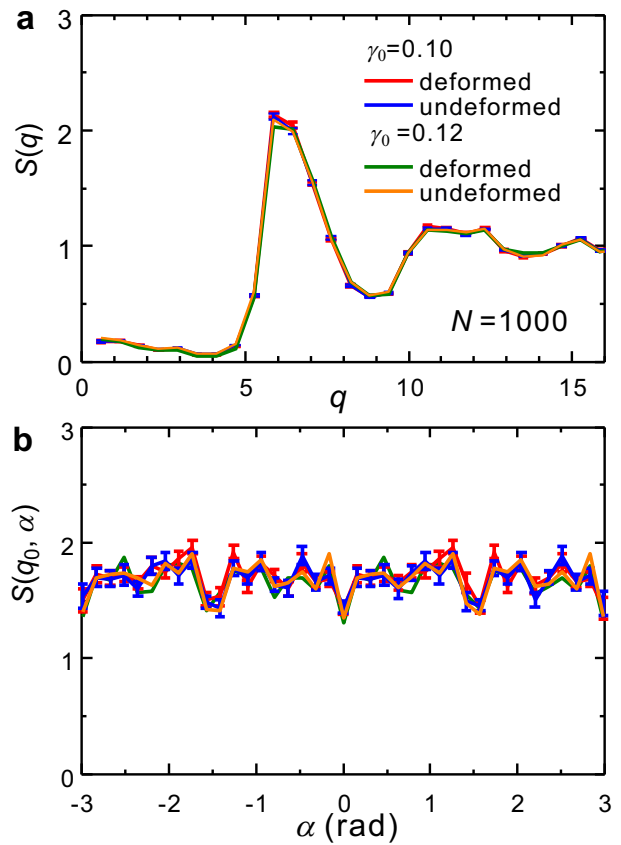,width=8.cm}
\caption{\label{fig4} 
Averaged static structure factors for `deformed' (when the stress is zero)
and undeformed (when the stress is maximal) states in the reversible 
phase at $\gamma_0=0.10$ and in the diffusive phase at $\gamma_0=0.12$, 
for $N=10^3$.
(a) Wavevector dependence of the structure with spherical average
over all directions.
(b) Angular dependence of the structure factor for $|\vec{q}|$ near the 
first peak as a function of the angle $\alpha$ defined the 
$(x,z)$ components of $\vec{q}$. 
Errorbars are shown for $\gamma_0=0.10$, to indicate that our 
relative resolution is very good (about 2 \%).}
\end{figure}

In Figure \ref{fig4}(a), we show the $q$-dependence of the averaged 
structure factor $S(q = |\vec{q}|)$ for these 4 cases 
and for $N=1000$. We observe that neither the particle reversibility 
nor the deformation seem to affect much the structure factor. 
We now resolve the angular dependence of $S(q_0,\alpha)$
using wavevectors in the $(x,z)$ plane having an amplitude 
close to the first peak in the range $q_0 = 5.9 - 7.0$. 
In Fig.~\ref{fig4}(b), we show that the $\alpha$-dependence of 
the structure factor for the 4 situations defined above 
is essentially inexistent. Importantly, we do not 
observe any difference for reversible and irreversible 
regimes across yielding. In particular we do not observe 
the oscillations that were detected in the experiments for the arrested
phase. 
Furthermore, we also checked that average values of other static quantities 
(such as the energy density and pair correlation functions) are 
similarly insensitive to the underlying dynamic transition.
These conclusions contrast with the results in Ref.~\cite{schall},
which are perhaps due to the larger shear rates employed in the experiment.
Another major contradiction with that work is our finding that yielding
does not correspond to the crossing point of $G'$ and $G''$.
Our results show that yielding is best interpreted as a loss 
of reversibility in the particle trajectories, which is a purely 
dynamical concept.  

\section{Conclusion}

Together, our results suggest that the yielding transition 
of jammed solids under large-amplitude oscillatory shear is 
accompanied by a first-order non-equilibrium phase transition,
rather than a continuous one. It marks the abrupt emergence 
of irreversible non-affine particle motion. 
The characteristic strain amplitudes 
obtained from rheology ($\gamma_{\times}$, $\gamma_{\rm pl}$)
and from microscopic dynamics ($\gamma_{\rm dyn}$, $\gamma_{\rm c}$)
are compiled in Fig.~\ref{fig1}(c). To a good approximation, 
we find that $\gamma_{\rm pl} \approx \gamma_{\rm dyn} \approx \gamma_{\rm c}$,
whereas $\gamma_{\times}$ is significantly larger, corresponding to a large 
amount of dissipated energy.  
The dynamic phase transition revealed by the discontinuous 
evolution of single-particle dynamics produces 
a smooth crossover in mechanical properties 
at a critical strain amplitude that appears unrelated to the 
crossing of $G'(\omega)$ and $G''(\omega)$. 
Our conclusions contrast with earlier 
claims of a continuous transition~\cite{ganapathy,foffi}, but appear 
in very good agreement with observations 
in a sheared emulsion \cite{luca}. We hope  our study 
will trigger further work in a broader variety of numerical and experimental
systems to fully establish its generality.

\acknowledgments
We thank L. Cipelletti, E. Tjhung, D. Truzzolillo for discussions, 
G. Szamel and S. Liese for exploratory simulations, and
T. Divoux and S. Manneville for exploratory experiments.
The research leading to these results has received funding
from the European Research Council under the European Union's Seventh
Framework Programme (FP7/2007-2013) / ERC Grant agreement No 306845
{ and funding from JSPS Kakenhi (No. 15H06263, 16H04025, 16H04034, and 16H06018).}

\end{document}